\documentclass[preprint,12pt]{elsarticle}

\usepackage{graphicx}

\usepackage{amssymb}

\usepackage{lineno}

\biboptions{comma,super,compress}

\usepackage[hidelinks]{hyperref}
\usepackage{amsmath}
\usepackage{bm}
\usepackage{listings}
\usepackage{xcolor}
\usepackage{booktabs}
\usepackage{float}

\definecolor{codegreen}{rgb}{0,0.6,0}
\definecolor{codegray}{rgb}{0.5,0.5,0.5}
\definecolor{codepurple}{rgb}{0.58,0,0.82}
\definecolor{backcolour}{rgb}{0.95,0.95,0.92}

\newcommand{\chemtrain}{{\tt chemtrain}}

\lstdefinestyle{mystyle}{
  backgroundcolor=\color{backcolour}, commentstyle=\color{codegreen},
  keywordstyle=\color{magenta},
  numberstyle=\tiny\color{codegray},
  stringstyle=\color{codepurple},
  basicstyle=\ttfamily\scriptsize,
  language=Python,
  morekeywords={True, False},
  frame=none,
  breakatwhitespace=false,         
  breaklines=true,                 
  captionpos=b,                    
  keepspaces=true,                 
  numbers=none,                    
  numbersep=5pt,                  
  showspaces=false,                
  showstringspaces=false,
  showtabs=false,                  
  tabsize=4
}
\newcounter{bla}

\journal{Computer Physics Communications}

\makeatletter
\def\ps@pprintTitle{%
 \def\@oddfoot{\parbox{\textwidth}{\em\sffamily\footnotesize \copyright{} 2025 The Authors. This manuscript version is made available under the CC-BY 4.0 license \url{https://creativecommons.org/licenses/by/4.0/}. Original DOI: \url{https://doi.org/10.1016/j.cpc.2025.109512}}}%
 \def\@evenfoot{\em\small Preprint \hfill \today}%
}
\makeatother

\begin{document}

\begin{frontmatter}



\title{{\chemtrain}: Learning Deep Potential Models via Automatic Differentiation and Statistical Physics}


\author[a]{Paul Fuchs \corref{first}}
\author[a,b]{Stephan Thaler \corref{first}}
\author[a]{Sebastien Röcken}
\author[a,c]{Julija Zavadlav \corref{author}}

\cortext[first] {Contributed equally}
\cortext[author] {Corresponding author (\textit{julija.zavadlav@tum.de})}

\address[a]{Multiscale Modeling of Fluid Materials, Department of Engineering Physics and Computation, TUM School of Engineering and Design, Technical University of Munich, Germany}
\address[b]{Valence Labs, Montreal, QC, Canada}
\address[c]{Atomistic Modeling Center, Munich Data Science Institute, Technical University of Munich, Germany}

\begin{abstract}
	
Neural Networks (NNs) are {effective} models for refining the accuracy
of molecular dynamics, opening up new fields of application.
Typically trained bottom-up, atomistic NN potential models can reach first-principle accuracy, while coarse-grained implicit solvent NN potentials surpass classical continuum solvent models.
However, overcoming the limitations of costly generation of accurate reference data and data inefficiency of common bottom-up training demands efficient incorporation of data from many sources.
This paper introduces the framework {\chemtrain} to learn sophisticated NN potential models through customizable training routines and advanced training algorithms.
These routines can combine multiple top-down and bottom-up algorithms, e.g., to incorporate both experimental and simulation data or pre-train potentials with less costly algorithms.
{\chemtrain} provides an object-oriented high-level interface to simplify the creation of custom routines. On the lower level, {\chemtrain} relies on JAX to compute gradients and scale the computations to use available resources.
We demonstrate the simplicity and importance of combining multiple algorithms in the examples of parametrizing an all-atomistic model of titanium and a coarse-grained implicit solvent model of alanine dipeptide.\\



\noindent {\bf PROGRAM SUMMARY}

\begin{small}
\noindent
{\em Program Title:} {\chemtrain}                                          \\
{\em CPC Library link to program files:} (to be added by Technical Editor) \\
{\em Developer's repository link:} \url{https://github.com/tummfm/chemtrain} \\
{\em Code Ocean capsule:} (to be added by Technical Editor)\\
{\em Licensing provisions:} Apache-2.0  \\
{\em Programming language:} python                                 \\
{\em Supplementary material:}                                 \\
{\em Nature of problem:}\\ 
Neural Network (NN) potentials provide the means to accurately model high-order many-body interactions between particles on a molecular level.
Through linear computational scaling with the system size, their high expressivity opens up new possibilities for efficiently modeling systems at a higher precision without resorting to expensive, finer-scale computational methods. 
However, as common for data-driven approaches, the success of NN potentials depends crucially on the availability of accurate training data.
Bottom-up trained state-of-the-art models can match ab initio computations closer than their actual accuracy but can still predict deviations from experimental measurements.
Including more accurate reference data can, in principle, resolve this issue, but generating sufficient data is infeasible even with less precise methods for increasingly larger systems.
Supplementing the training procedure with more data-efficient methods can limit required training data\cite{1}. In addition, the models can be fully or partially trained on macroscopic reference data\cite{2,3}.
Therefore, a framework supporting a combination of multiple training algorithms could further expedite the success of NN potential models in various disciplines.\\
{\em Solution method:}\\
We propose a framework that enables the development of NN potential models through customizable training routines. The framework provides the top-down algorithm Differentiable Trajectory Reweighting\cite{2} and the bottom-up learning algorithms Force Matching\cite{1} and Relative Entropy Minimization\cite{1}.
A high-level object-oriented API simplifies combining multiple algorithms and setting up sophisticated training routines such as active learning.
At a modularly structured lower level, the framework follows a functional programming paradigm relying on the machine learning framework JAX\cite{4} to simplify the creation of algorithms from standard building blocks, e.g., by deriving microscopic quantities such as forces and virials from any JAX-compatible NN potential model and scaling computations to use available resources.\\

\end{small}
\end{abstract}

\end{frontmatter}

\section{Introduction}
\label{section:introduction}

Molecular Dynamics (MD) is an essential tool in multiple disciplines, ranging from computational biology to mechanical engineering \cite{hartmannMechanismCollagenFolding2021, liMolecularDynamicsPredictions2011, fatriansyahMolecularDynamicsSimulation2019}.
Classical force fields\cite{cornellSecondGenerationForce1995, nikitinNewAMBERcompatibleForce2014, marrinkCoarseGrainedModel2004} concurrently developed with established MD software\cite{thompsonLAMMPSFlexibleSimulation2022, abrahamGROMACSHighPerformance2015, andersonGeneralPurposeMolecular2008,andersonHOOMDbluePythonPackage2020} offer inexpensive but accurate modeling of molecular systems in specialized fields.
However, the simplicity of classical force fields prohibits modeling quantum effects such as charge transfer and bond-breaking.
At the same time, available computational resources limit the possible system size and simulation time even for very simple classical atomistic force fields.
Therefore, multiscale modeling approaches are crucial to extending MD simulation to systems with large scale separation\cite{tozziniMultiscaleModelingProteins2010, zavadlavAdaptiveResolutionSimulation2014, zavadlavOrderInteractionsDNA2017, thalerBackmappingAugmentedAdaptive2020}.

Recently, Neural Networks (NNs) potential models emerged as a flexible approach to supplement or replace classical multiscale simulation techniques.
NN potential architectures\cite{behlerConstructingHighdimensionalNeural2015,smithANI1ExtensibleNeural2017,schuttSchNetContinuousfilterConvolutional2017, gasteigerDirectionalMessagePassing2022, gasteigerFastUncertaintyAwareDirectional2022, batznerEquivariantGraphNeural2022, musaelianLearningLocalEquivariant2023} model high-order multi-body interactions using pre-defined\cite{behlerAtomcenteredSymmetryFunctions2011, smithANI1ExtensibleNeural2017} or learned\cite{schuttSchNetContinuousfilterConvolutional2017, gasteigerDirectionalMessagePassing2022, gasteigerFastUncertaintyAwareDirectional2022, batznerEquivariantGraphNeural2022, musaelianLearningLocalEquivariant2023} features of the local environment of the atoms.
Therefore, the computational effort of employing NN-based force fields scales linearly with the size of the system, similar to classical force fields.
At the atomistic scale, the NN models can predict energies and forces with a precision similar to first-principle methods\cite{smithANI1ExtensibleNeural2017, smithApproachingCoupledCluster2019, gasteigerDirectionalMessagePassing2022}.
Consequently, NN-enhanced MD can generate trajectories with similar accuracy but is less computationally expensive than ab initio MD\cite{iftimieInitioMolecularDynamics2005} and introduces {electronic} quantum effects to the whole system, unlike hybrid quantum-mechanical / molecular-mechanical modeling\cite{tzeliouReviewQMMM2022}.
In coarse-grained (CG) MD approaches, NN potentials can approximate important many-body interactions of the potential of mean force often neglected in classical CG potential models\cite{zaporozhetsMultibodyTermsProtein2023}.
Therefore, NN potentials also enable highly accurate mesoscopic modeling, surpassing, e.g., classical implicit solvation models\cite{chenMachineLearningImplicit2021, thalerDeepCoarsegrainedPotentials2022}.

NN potential models are data-driven approaches for which optimal para\-metri\-zations must be inferred from experiments or simulations.
The parameterization of potential models is typically categorized into top-down and bottom-up strategies\cite{noidPerspectiveCoarsegrainedModels2013}.
Bottom-up approaches\cite{shellRelativeEntropyFundamental2008, ercolessiInteratomicPotentialsFirstPrinciples1994, noidMultiscaleCoarsegrainingMethod2008} learn parameters by fitting predictions of the model to large amounts of high-resolution data from a finer-scaled reference model.
Top-down approaches\cite{lyubartsevCalculationEffectiveInteraction1995, reithDerivingEffectiveMesoscale2003, norgaardExperimentalParameterizationEnergy2008, whiteEfficientMinimalMethod2014, whiteDesigningFreeEnergy2015} adjust the model such that simulations reproduce experimentally determined macroscopic properties.
The majority of software packages supporting the training of NN-based potential models focus on bottom-up strategies\cite{ruppGuestEditorialSpecial2024, wangDeePMDkitDeepLearning2018, zengDeePMDkitV2Software2023, schuttSchNetPackDeepLearning2019, schuttSchNetPackNeuralNetwork2023, barrettGPUAcceleratedMachineLearning2019, khorshidiAmpModularApproach2016, shuaibi_2023_8151492, pleFeNNolEfficientFlexible2024, artrithImplementationArtificialNeuralnetwork2016, cooperEfficientTrainingANN2020, singraberLibraryBasedLAMMPSImplementation2019, kolbDiscoveringChargeDensity2017, Ramsundar-et-al-2019, fanGPUMDPackageConstructing2022, songPhysNetMeetsCHARMM2023, huangSPONGEGPUAcceleratedMolecular2022} by matching potential energies, forces, and other microscopic quantities via the Force Matching (FM) algorithm\cite{ercolessiInteratomicPotentialsFirstPrinciples1994}.
With sufficient reference data, bottom-up trained state-of-the-art potential can match the accuracy of the data generating (e.g., density functional theory) method \cite{smithANI1ExtensibleNeural2017, smithApproachingCoupledCluster2019, gasteigerDirectionalMessagePassing2022}.
Nevertheless, the models cannot surpass the method's accuracy, affecting the quality of macroscopic predictions\cite{rockenPredictingSolvationFree2024, frohlkingEmpiricalForceFields2020}.
Including reference data from more precise (e.g., coupled cluster) methods can alleviate this issue\cite{smithApproachingCoupledCluster2019}.
Still, the investigation of increasingly larger systems, particularly when developing CG models, even renders the generation of sufficient data with less precise methods computationally infeasible.
One option is to limit required reference data by supplementing the training procedure with more data-efficient, albeit more expensive, methods such as Relative Entropy Minimization (RM) for CG systems\cite{thalerDeepCoarsegrainedPotentials2022}.
More generally, a viable alternative to enforce consistency with macroscopic observables for models at all scales is to partially\cite{rockenAccurateMachineLearning2024,rockenPredictingSolvationFree2024} or fully\cite{thalerLearningNeuralNetwork2021,navarroTopDownMachineLearning2023} employ top-down learning.
Classically, top-down learning approaches\cite{reithDerivingEffectiveMesoscale2003, lyubartsevCalculationEffectiveInteraction1995, whiteEfficientMinimalMethod2014, whiteDesigningFreeEnergy2015} are focused on training potential models with particular functional forms, but can be generalized to support a wider range of macroscopic observables and potential models through reweighting techniques\cite{norgaardExperimentalParameterizationEnergy2008, liIterativeOptimizationMolecular2011, carmichaelNewMultiscaleAlgorithm2012, wangSystematicParametrizationPolarizable2013, thalerLearningNeuralNetwork2021, wangDMFFOpenSourceAutomatic2023}.
Particularly, the Differentiable Trajectory Reweighting (DiffTRe) algorithm\cite{thalerLearningNeuralNetwork2021} generalizes reweighting based top-down learning to NN potential models by leveraging Automatic Differentiation (AD) and differentiable physics MD packages\cite{barrettGPUAcceleratedMachineLearning2019, doerrTorchMDDeepLearning2021, schoenholzJAXFrameworkDifferentiable2021}, available for many established machine learning libraries\cite{abadiTensorFlowLargeScaleMachine, paszkePyTorchImperativeStyle2019, frostigCompilingMachineLearning}. Additionally, the DiffTRe algorithm enables the implementation of reweighting-based RM as a more data-efficient alternative or complement to FM to learn CG NN potential models.
However, available software\cite{wangDMFFOpenSourceAutomatic2023, barrettGPUAcceleratedMachineLearning2019} only provide a DiffTRe-like method with support for classical force fields or adjust NN potential models through minimally-biasing methods\cite{whiteEfficientMinimalMethod2014, whiteDesigningFreeEnergy2015}, which correct macroscopic properties by adding a minimal correction to the NN potential model, but are not suitable for learning transferable potential models\cite{frohlkingEmpiricalForceFields2020}.

In this paper, we propose {\chemtrain}, a framework that enables the development of sophisticated NN potentials through highly customizable training routines.
As elementary algorithms, {\chemtrain} provides the well-known bottom-up approach of FM\cite{thalerDeepCoarsegrainedPotentials2022} and its uncertainty-aware probabilistic extension\cite{thalerScalableBayesianUncertainty2023} trough an interface to JaxSGMC\cite{thalerJaxSGMCModularStochastic2024}, the bottom-up approach of RM\cite{thalerDeepCoarsegrainedPotentials2022} and the top-down approach DiffTRe\cite{thalerLearningNeuralNetwork2021}.
An object-oriented high-level Application Programming Interface (API) simplifies the creation of custom training procedures, e.g., by combining algorithms\cite{rockenAccurateMachineLearning2024} or employing active learning strategies\cite{thalerActiveLearningGraph2024}.
At the lower level, {\chemtrain} is modularly structured, following a functional paradigm to leverage JAX\cite{frostigCompilingMachineLearning} transformations to simplify setting up algorithms from basic building blocks.
These transformations enable, e.g., the automatic derivation of microscopic quantities, such as forces and virials, from any JAX-compatible NN potential models and the parallelization of computations across multiple devices to efficiently use available resources.
Hence, {\chemtrain} aims to provide a simple, flexible, and computationally efficient software solution to develop sophisticated potential models for various fields.

This paper is structured as follows:
To bridge between users with machine learning and physics backgrounds, we provide a reiteration of the fundamentals of MD in section~\ref{section:theory} and the foundation of top-down/bottom-up training algorithms in section~\ref{sec:machine_learning}, respectively.
In section~\ref{section:software}, we explain the design of {\chemtrain} and provide a detailed overview of the structure, including a summary of the most important building blocks.
Section~\ref{section:examples} outlines the importance and usage of {\chemtrain} at two relevant examples combining training algorithms to efficiently learn improved CG and atomistic potential models.
Finally, we present possibilities for the future of {\chemtrain} in section~\ref{sec:conclusion}.

\section{Molecular Dynamics and Statistical Mechanics}
\label{section:theory}
\label{subsec:md_basis}

In the following section, we briefly reiterate the fundamentals of MD\cite{tuckermanStatisticalMechanicsTheory2015} as a computational tool to model the time evolution of a molecular system.
Under the assumption of ergodicity, we outline a connection between the microscopic long-term simulation of a single system and the macroscopic probabilistic statistical mechanics perspective on many identical but independent systems.
This connection forms a basis for the variational approaches to top-down and bottom-up learning of potential models in section~\ref{sec:machine_learning}.

MD aims to simulate material properties at a molecular level by solving the classical equations of motion for a system of point-mass-like particles\cite{tuckermanStatisticalMechanicsTheory2015}.
For a system of $N$ particles with positions $\bm r = (\bm r_1, \ldots, \bm r_N) \in \mathbb{R}^{3N}$ and momenta $\bm p = (\bm p_1, \ldots, \bm p_N) \in \mathbb{R}^{3N}$, Newtons second law describes the evolution of the system via $6N$ coupled differential equations
\begin{linenomath}\begin{align}
     \dot{\bm r_i} = \frac{\bm p_i}{m_i}, \quad \dot{\bm p_i} = \bm F_i(\bm r, \bm p_i),
\end{align}\end{linenomath}
where $m_i$ denotes the mass of the $i$-th particle.
The forces $\bm F$ model all interactions between the particles and can even account for non-conservative interactions with the environment.
However, the dynamics of an isolated piece of matter must obey to the conservation of energy.
Therefore, one typically restricts the interaction forces to be conservative by defining a scalar potential $U(\bm r)$ such that
\begin{linenomath}\begin{align}
    \bm F_i(\bm r) = - \frac{\partial U(\bm r)}{\partial \bm r_i}.\label{eq:conservative_force}
\end{align}\end{linenomath}
Combined with the kinetic energy $K$ for simple point masses
\begin{linenomath}\begin{align}
  K(\bm p) = \sum_{i=1}^N \frac{|\bm p_i|^2}{m_i},
\end{align}\end{linenomath}
the scalar potential forms the Hamiltonian $ H(\bm r, \bm p) =  K(\bm p) +  U(\bm r)$ equal to the total energy of the system.
Many sophisticated numerical schemes exist for solving the ordinary Newtonian or Hamiltonian equations.
However, integrating the Hamiltonian equations with a symplectic integrator bounds the error of energy conservation and enables stable simulations of molecular systems over very long time spans.

MD models matter at a microscopic level but aims to reproduce and understand experiments at macroscopic scales.
Therefore, it is necessary to interpret the microscopic results from the macroscopic perspective of statistical mechanics and thermodynamics\cite{tuckermanStatisticalMechanicsTheory2015}.
The basis of statistical mechanics relies on the concept of ensembles.
An ensemble $\mathcal E$ is a collection of all the identical systems that reproduce a set of macroscopic observables.
At time $t$, the microscopic states $\bm x_\lambda$ of the systems $\lambda \in \mathcal E$ are distributed in phase space according to a distribution function $p(\bm x, t)$.
Therefore, a macroscopic observable $A$ obtained as the average of instantaneous microscopic properties $a(\bm x_\lambda)$ for all ensemble members $\lambda \in \mathcal E$ is equivalently given as an expectation on the ensemble distribution function
\begin{linenomath}\begin{align}
A(t) = \frac{1}{|\mathcal E|}\sum_{\lambda \in \mathcal E} a(\bm x_\lambda) = \langle a(\bm x)\rangle_{p(\bm x, t)} = \int a(\bm x)p(\bm x, t)d\bm x.
\end{align}\end{linenomath}

Since the microscopic states of the systems evolve according to the same dynamics $\dot{\bm x} = \xi(\bm x)$, the ensemble distribution function $p(\bm x, t)$ is directly linked to the dynamics\cite{tuckermanClassicalStatisticalMechanics1999}.
However, practically exploiting this link through MD relies on the assumption of ergodicity and thermodynamic equilibrium $p(\bm x, t) = p(\bm x)$.
Generally, two systems in an ensemble might reside in separate regions of phase space and never visit the same states.
However, a system in an ergodic ensemble will evolve through all points in phase space with a frequency proportional to $p(\bm x)$ if given an infinite amount of time.
Therefore, the long-term average over the evolution of a such single system
\begin{linenomath}\begin{align}
    A = \langle a(\bm x) \rangle_\tau = \lim_{t \rightarrow \infty}\frac{1}{t}\int_0^t a(\bm x(\tau))d\tau\label{eq:time_average}
\end{align}\end{linenomath}
approaches the phase-space average over all accessible microstates.
Hence, a single MD simulation suffices to predict the equilibrium observables of an ergodic ensemble.

A commonly employed ensemble for MD is the canonical ensemble of $N$ particles in a space of constant volume, maintaining an average temperature $T$ via thermal exchange with an environment.
In the canonical ensemble, the position variables $\bm r$ are independent of the momenta $\bm p$, leading to a reduced form of the equilibrium distribution
\begin{linenomath}
\begin{align}
    p_U(\bm r) = e^{-\beta U(\bm r) - \beta Q_U}, \quad Q_U = -\frac{1}{\beta}\log\int e^{-\beta  U(\bm r)}d\bm r,\label{eq:canonical_density}
\end{align}
\end{linenomath}
where $\beta = \frac{1}{k_B T}$ is the inverse of the microscopic temperature $k_B T$ and $k_B$ is the Boltzmann constant.
While statistical mechanics arrive at $p(\bm x)$ by marginalizing out degrees of freedom of an excessively larger thermal bath, simulation of these degrees of freedom is inconvenient.
Instead, in practical MD one perturbs the Hamiltonian dynamics of the system, e.g., by adding a chained Nosé-Hoover thermostat\cite{martynaNoseHooverChains1992} or performing stochastic Langevin Dynamics \cite{bussiAccurateSamplingUsing2007} such that the new dynamics $\xi(\bm x)$ lead to a quasi-canonical equilibrium distribution.

\section{Machine Learning of Potential Models}
\label{sec:machine_learning}

\subsection{Top-Down Learning}
\label{subsec:top_down}

Top-down learning aims to find a potential model $U$ such that the molecular system reproduces a set of macroscopic targets $\{\hat A_k\}$.
In a variational formulation, the optimal potential parameters $\theta$ minimize the difference $\Delta_k = \hat{\mathcal A}_k - \mathcal A_k[U_\theta]$ between the target values $\hat{\mathcal A}_k$ and simulated predictions $\mathcal A_k[U_\theta]$ trough a weighted sum of errors
\begin{linenomath}
    \begin{align}
    \mathcal L(\theta) = \sum_{k=1}^K \gamma_k \ell_k(\Delta_k)\label{eq:difftre_loss},
    \end{align}
\end{linenomath}
where $\gamma_k$ weights per-target contributions based on, e.g., the mean-squared error $\ell(\Delta_k) = \frac{1}{2\dim \Delta_k}\lVert\Delta_k\rVert^2$.
We introduced $\mathcal A$ to conveniently account for macroscopic observables consisting of multiple ensemble averages, such as the heat capacity\cite{sadusMolecularSimulationOrthobaric2019}.

Gradient-based optimization is a widely used and efficient method of training NN models.
At first glance, implementing the computation of $\mathcal A_K[U_\theta]$ in a differentiable MD framework provides the necessary gradients of the loss via AD.  
However, molecular systems often exhibit chaotic dynamics\cite{braxenthalerChaosProteinDynamics1997}, such that small perturbations of the system can critically affect the solution $\bm x(t)$ of the differential equations.
Consequently, employing AD to differentiable simulations does not guarantee meaningful gradients\cite{metzGradientsAreNot2022}, which are even susceptible to the exploding gradients issue\cite{ingraham2019, thalerLearningNeuralNetwork2021} and costly for many integration steps.
Therefore, MD simulations are suitable to efficiently predict the value of the loss but can be insufficient for gradient-based optimization\cite{hanRefiningPotentialEnergy2024, thalerLearningNeuralNetwork2021}.

From the probabilistic perspective of thermodynamic perturbation theory\cite{zwanzig2004}, a change in potential from $\tilde U$ to $U = \tilde U - \Delta U$ affects the relative probability $p_U/p_{\tilde U}$ of conformations $\bm r$.
Interpreting this relative probability as weighting function $w_U = p_U/p_{\tilde U}$ establishes a connection between the ensemble averages of the perturbed and unperturbed system
\begin{linenomath}\begin{align}
    A_U = \langle a(\bm r) \rangle_{p_U} = \int a(\bm r)w_U(\bm r)p_{\tilde U}(\bm r)d\bm r = \langle w_U(\bm r) a(\bm r) \rangle_{p_{\tilde U}},\label{eq:perturbation_formula}
\end{align}\end{linenomath}
as long as both distribution functions overlap.
The canonical ensemble fulfills the overlap criterion for any finite perturbation $\Delta U$ with a corresponding weight function 
\begin{linenomath}\begin{align}
    w_U(\bm r) = \frac{e^{-\beta\Delta U(\bm r)}}{\left\langle e^{-\beta \Delta U(\bm r)}\right\rangle_{p_{\tilde U}}} \label{eq:weight_function}.
\end{align}\end{linenomath}
Assuming that a parameter dependence of $U$ is introduced only through $\Delta U_\theta$, the weighted ensemble average leads to a well-defined expression for the observable gradients
\begin{linenomath}
    \begin{align}
   \frac{\partial A_U}{\partial \theta} = \left\langle \frac{\partial a(\theta, \bm r)}{\partial \theta}w_U(\theta, \bm r) + a(\theta, \bm r) \frac{\partial w_U(\theta, \bm r)}{\partial \theta} \right\rangle_{p_{\tilde U}}.\label{eq:perturbation_gradients}
    \end{align}
\end{linenomath}
The direct application of thermodynamic perturbation theory is often analytically intractable as it requires the evaluation of ensemble averages given in equations~\eqref{eq:perturbation_formula} and \eqref{eq:perturbation_gradients}.
However, combined with Monte Carlo (MC) or MD as in the Umbrella sampling method\cite{torrie1977}, perturbation theory provides a basis for viable approaches to top-down learning\cite{norgaardExperimentalParameterizationEnergy2008, liIterativeOptimizationMolecular2011, thalerLearningNeuralNetwork2021, wangDMFFOpenSourceAutomatic2023}.

One particularly flexible approach is the recently proposed DiffTRe algorithm\cite{thalerLearningNeuralNetwork2021}.
Starting the optimization from an initial guess $U^{(0)}$, the algorithm draws decorrelated samples $\{\bm r^{(i)}\}_{i=1}^D$ from $\tilde U \leftarrow U^{(0)}$ via MD simulations.
Relying on a fully-differentiable implementation of $a_\theta$, $\Delta U_\theta$, and the reweighting procedure, the algorithm leverages AD to update $\theta$ based on the gradients of the loss~\eqref{eq:difftre_loss}.
To account for the statistical error from approximating the exact perturbation formula~\eqref{eq:perturbation_formula} with a finite number of samples\cite{ceriotti2012}, the algorithm measures the effective sample size\cite{carmichaelNewMultiscaleAlgorithm2012}
\begin{linenomath}
    \begin{align}
    \operatorname{ESS} = e^{-\sum_{i=1}^D w_U^{(i)}\log w_U^{(i)}} \in [1, D]
    \end{align}
\end{linenomath}
and resamples new reference states from the current potential candidate $\tilde U \leftarrow U^{(i)}$ whenever the ESS drops below a pre-defined threshold.
Therefore, the DiffTRe algorithm enables training expressive potential models with complicated functional forms, such as NN potential models, based on equilibrium macroscopic observables.
Additionally, an extension to dynamic properties is, in principle, also possible\cite{hanRefiningPotentialEnergy2024}.

\subsection{Bottom-Up Learning}
\label{subsec:bottom_up}

We approach the problem of bottom-up learning from the perspective of coarse-graining from a fine-grained (FG) to a coarse-grained (CG) particle system.
A typical example is modeling an FG atomistic representation with a coarser approximation, in which several atoms are grouped into effective beads modeled as single particles.
However, learning effective atomistic models from ab initio data also complies with this idea \cite{izvekovMultiscaleCoarseGrainingMethod2005}.

A common first step in coarse-graining is to link the CG system to its FG reference by selecting a mapping $\mathcal M: \mathbf{R}^{3n} \rightarrow \mathbf{R}^{3N}$.
This mapping relates the positions of the $N$ CG particles $\bm R$ to the positions of the $n$ fine-grained particles $\bm r$ via
\begin{linenomath}\begin{align}
    \bm R = \mathcal M(\bm r).
\end{align}\end{linenomath}
If the FG and CG models act on the same scale, e.g., in training an atomistic model on ab initio reference data, the identity function is a trivial choice for the mapping.
In coarse-graining, a computationally convenient choice of mapping places the position of the CG particle at the center of mass of the fine-grained particles that constitute an effective bead\cite{izvekovMultiscaleCoarseGrainingMethod2005}.
However, more sophisticated mappings, e.g., based on dimensionality reducing NNs\cite{wangCoarsegrainingAutoencodersMolecular2019} can also be useful.

\subsubsection{Force Matching}
\label{subsubsec:force_matching}

The FM approach was originally proposed as an efficient method to learn atomistic potentials from large amounts of first-principle data\cite{ercolessiInteratomicPotentialsFirstPrinciples1994}.
However, it is also a practical and theoretically well-founded approach to couple particle-based models of different scales\cite{izvekovMultiscaleCoarseGrainingMethod2005,noidMultiscaleCoarsegrainingMethod2008}.

The motivation for employing FM as a coarse-graining approach lies in the uniqueness of forces of a thermodynamic consistent model\cite{noidMultiscaleCoarsegrainingMethod2008}.
A CG model $U^\text{CG}$ is consistent with the reference model $U^\text{FG}$ and the mapping $\mathcal M$, if it generates CG conformations $\bm R \sim p^\text{CG}(\bm R)$ that match the mapped samples $\mathcal M(\bm r)$ from the fine-grained reference $\bm r \sim p^\text{FG}(\bm r)$.
More formally, the distribution $p^\text{CG}$ of the consistent model $U^\text{CG}$ is
\begin{linenomath}\begin{align}
    p^\text{CG}(\bm R) = %
    \left\langle \delta\left[\mathcal M(\bm r) - \bm R\right]\right\rangle_{p^\text{FG}}\label{eq:consistency}.
\end{align}\end{linenomath}
From equations~\eqref{eq:conservative_force}~and~\eqref{eq:canonical_density} follows that the consistent distribution uniquely defines the forces $\bm F$ acting on the CG sites $\bm R$ as
\begin{linenomath}\begin{align}
    \bm F_I(\bm R) & = - \frac{1}{\beta}\frac{\partial}{\partial \bm R_I} \log p^\text{CG}(\bm R).
\end{align}\end{linenomath}
Under the assumption of a linear mapping $\mathcal M(\bm R) = C\bm R$ and a linear mapping of forces $\mathcal F(\bm f) = D\bm f$, both with weak constraints, it is possible to prove that these forces are the unique minimum of the functional\cite{noidMultiscaleCoarsegrainingMethod2008}
\begin{linenomath}\begin{align}
    \chi_{\bm F}^2[\bm F] = \frac{1}{3N}\sum_{I=1}^N \left\langle\left\lVert \mathcal F_I(\bm f(\bm r)) - \bm F_I(\mathcal M(\bm r)) \right\rVert^2\right\rangle_{p^\text{FG}}\label{eq:FM_residual}.
\end{align}\end{linenomath}
Therefore, minimizing the functional for a parameterized model $\tilde{\bm F_\theta} = -\frac{\partial U_\theta}{\partial \bm R_I}$ provides a well-based variational approach to coarse-graining.

An additional step is necessary to apply FM practically.
Analogous to top-down coarse-graining, computing the exact value of the residual in equation~\eqref{eq:FM_residual} is analytically intractable.
Instead, one must resort to estimates based on a finite number of samples $\mathcal D = \left\{ \left(\mathcal F^{(i)}, \bm R^{(i)}\right)\right\}_{i=1}^D$ from the FG reference system, e.g., via MC or MD simulations.
Then, it is possible to employ classical numerical optimization procedures\cite{ercolessiInteratomicPotentialsFirstPrinciples1994, noidMultiscaleCoarsegrainingMethod2008a} or stochastic gradient-based methods\cite{kingma2017} to minimize the approximate loss function 
\begin{linenomath}\begin{align}
    \mathcal L_{\bm F}(\theta) = \frac{1}{3ND}\sum_{i=1}^D\sum_{I=1}^N\left\lVert  \mathcal{F}_{I}^{(i)} - \bm F_I\left(\theta, \bm R^{(i)}\right)\right\rVert^2.\label{eq:FM_loss}
\end{align}\end{linenomath}

For linear models with infinite data, a unique optimal model closest to the consistent model exists\cite{noidMultiscaleCoarsegrainingMethod2008}.
However, training more complex models on inadequately sampled data introduces uncertainty with possibly severe consequences on the robustness of the learned models\cite{stockerHowRobustAre2022}.
By employing uncertainty-aware inference methods, it is possible to account for the effects of finite data size and model pluralism. 
For example, adopting the FM loss of equation~\eqref{eq:FM_loss} as a likelihood into a probabilistic formulation
\begin{linenomath}\begin{align}
   p(\mathcal \theta | \mathcal D) \propto p(\theta) \prod_{k=1}^D \mathcal{N}\left(\mathbf F_\theta\left(\bm R^{(k)}\right) \mid \mathcal F^{(k)}, \sigma^2 \mathbb I\right)
\end{align}\end{linenomath}
enables the application of Bayesian Inference.
However, especially for neural network models, exact Bayesian Inference, e.g., via stochastic-gradient MC\cite{wellingBayesianLearningStochastic2011} is costly\cite{thalerScalableBayesianUncertainty2023}.
Hence, approximate or non-Bayesian schemes such as Drop-out MC\cite{wenUncertaintyQuantificationMolecular2020} or the Deep Ensemble method\cite{musilFastAccurateUncertainty2019, thalerScalableBayesianUncertainty2023} are popular alternatives.

Training only a single but accurate model is desirable for most practical applications.
At the atomistic scale, additional training on other potential-dependent properties\cite{ercolessiInteratomicPotentialsFirstPrinciples1994} increases available information and improves accurate predictions of these properties.
Typical examples are potential energy, which provides information about the relative probability of samples compared to the exclusive gradient information from forces, and virials, which are important, e.g., for NPT simulations.
In principle, matching these additional quantities is also possible for FM in the CG setting\cite{dasMultiscaleCoarsegrainingMethod2010}.
However, thermodynamic observables at the atomistic level are generally not directly applicable at the CG scale\cite{wagnerRepresentabilityProblemPhysical2016}, complicating a simple extension to arbitrary quantities.
Additionally, the extended approach enables the training of models that couple the prediction of forces for MD with the prediction of non-classical properties, e.g., electronic properties\cite{unkePhysNetNeuralNetwork2019}.
Hence, a more general formulation of the FM approach at the atomistic scale matches additional parameter-dependent quantities $a_k$ to targets $\hat a_k$ by minimizing a weighted sum of losses
\begin{linenomath}\begin{align}
    \mathcal L(\theta) = \gamma_{\bm F}\mathcal L_{\bm F}(\theta) +  \frac{1}{D}\sum_{i=1}^D\sum_{k=1}^K \gamma_k\left\lVert \hat a_k^{(k)} - a_k\left(\theta, \bm R^{(k)}\right)\right\rVert^2,
\end{align}\end{linenomath}
where $\gamma_k$ weights the contributions of the per-target losses, e.g., $\mathcal L_{\bm F}$, to the total loss.

\subsubsection{Relative Entropy Minimization}

The relative entropy, also known as Kullback-Leibler divergence\cite{kullback1951}, is a classical information-theoretical measure of the similarity of two distributions.
Descriptively, minimizing the relative entropy maximizes the likelihood of obtaining a sample from a target distribution when drawn from an independent reference distribution\cite{shellRelativeEntropyFundamental2008}.
Hence, RM provides an alternate foundation for learning good approximations of molecular systems.
For a reference distribution $p^\text{CG}(\bm r)$ from a surrogate model, the relative entropy towards a target distribution $p^\text{FG}(\bm r)$ is
\begin{linenomath}\begin{align}
    S_\text{rel} = \int p^\text{FG}(\bm r)\log\frac{p^\text{FG}(\bm r)}{p^\text{CG}(\bm r)}d\bm r \geq 0.\label{eq:rel_entropy}
\end{align}\end{linenomath}
In the context of coarse-graining, the parametrizable model $p_\theta^\text{CG}(\bm R)$ does not act on the fine-grained sites $\bm r$ but on the CG sites $\bm R$ connected through the mapping $\mathcal M$.
Without any prior assumption, any FG conformation $\bm r$ complying with a CG conformation $\bm R$ is equally likely, i.e., $p(\bm r \mid \bm R) \propto \delta(\mathcal M(\bm R) - \bm r)$.
Thus, by the law of total probability, the CG model defines the probability of FG conformations as
\begin{linenomath}\begin{align}
    p^\text{CG}(\bm r) = \int p(\bm r \mid \bm R)p_U^\text{CG}(\bm R)d\bm R = \frac{p^\text{CG}_U(\mathcal M(\bm r))}{\Omega_{\mathcal M}(\mathcal M(\bm r))}\label{eq:inv_mapping_cond_prob},
\end{align}\end{linenomath}
where $\Omega$ ensures the proper normalization of the conditional probability.
Therefore, the relative entropy~\eqref{eq:rel_entropy} between a canonical FG and canonical CG system is
\begin{linenomath}\begin{align}
    S_\text{rel} = \beta \left\langle U^\text{CG}(\mathcal M(\bm r)) - U^\text{FG}(\bm r) \right\rangle_\text{FG} - \beta(Q^\text{CG}_U - Q^\text{FG}) + S_\text{map},
\end{align}\end{linenomath}
with an irreducible contribution $S_\text{map} = \langle \log \Omega_{\mathcal M}(\mathcal M(\bm r))\rangle_{p^\text{FG}}$ arising from the degeneracy of the mapping.

In general, the relative entropy is hard to estimate since the mapping entropy $S_\text{map}$, and the free energies $Q^\text{FG}$ and $Q^\text{CG}_U$ are not simple ensemble averages.
However, $S_\text{map}$ and $Q^\text{FG}$ are irrelevant for gradient-based optimization since they depend only on the mapping or the FG reference distribution, respectively, but are constant with respect to the parameters $\theta$.
Analogously, the absolute free energy $Q_\theta^\text{CG}$ has the same gradients as a free energy difference $\Delta Q_\theta^\text{CG}$ with respect to a fixed reference potential.
Unlike for the absolute free energy, free energy perturbation methods\cite{carmichaelNewMultiscaleAlgorithm2012, zwanzig2004} enable the estimation of $\Delta Q_\theta^\text{CG}$ trough ensemble averages.
Therefore, the alternative loss function
\begin{linenomath}\begin{align}
    \mathcal L_\text{RE}(\theta) = \beta \left\langle U_\theta^\text{CG}(\mathcal M(\bm r))\right\rangle_\text{FG} - \beta \Delta A_\theta^\text{CG}\label{eq:RE_loss}
\end{align}\end{linenomath}
consists only of ensemble averages and differs from the relative entropy only by a constant.
Hence, minimizing $\mathcal L$ with respect to $\theta$ is equivalent to minimizing the relative entropy $S_\text{rel}$.
In a practical implementation, the first summand of the loss can be estimated based on a finite number of samples from the FG system, e.g., like in the FM algorithm.
For the second term, the DiffTRe algorithm provides a fully differentiable approach to compute perturbation-based estimates of the free energy difference\cite{rockenPredictingSolvationFree2024}.

\subsection{\texorpdfstring{$\Delta$}{Delta}-Learning Approach}

Adequate sampling of the conformational space is essential to the success of the presented learning methods.
However, data obtained through MD simulations is biased, resulting in infrequently sampling of some regions.
For example, strong repulsive forces at short intermolecular distances encode the Pauli exclusion principle but cause infrequent sampling of conformations with close particles.
From only a few of these samples, it is difficult to learn these short-range forces via bottom-up learning methods\cite{dasMultiscaleCoarsegrainingMethod2009}.
Analogously, top-down learning via the DiffTRe method requires sufficient overlap between the distributions at subsequent reweighting steps.
Consequently, sampling irrelevant conformations from an initial potential that does not model crucial properties of the interactions can severely impede learning success. 
Hence, a popular choice for the potential model in classical coarse-graining\cite{soperEmpiricalPotentialMonte1996, dasMultiscaleCoarsegrainingMethod2009} and neural-networks-based learning\cite{wangMachineLearningCoarseGrained2019, thalerLearningNeuralNetwork2021, thalerDeepCoarsegrainedPotentials2022} includes known properties through an additive prior potential
\begin{linenomath}\begin{align}
    U(\theta, \bm r) = \Delta U(\theta, \bm r) + U^\text{Prior}(\bm r).
\end{align}\end{linenomath}
The parametric model only learns the difference to the prior potential, leading to the $\Delta$-learning terminology \cite{Ramakrishnan2015}.
Especially for learning NN potential models, classical atomistic or CG force fields and functional forms provide a computational affordable basis for choosing a suitable prior potential\cite{chenMachineLearningImplicit2021, thalerLearningNeuralNetwork2021, thalerDeepCoarsegrainedPotentials2022}.

\section{Software}
\label{section:software}

We developed {\chemtrain} to advance the application of powerful potential models by facilitating the training of models via popular algorithms and promoting the development of new and elaborate training strategies. 
We structured {\chemtrain} into two layers to accomplish both objectives: At the lower level, we designed {\chemtrain} following a functional paradigm, using modular building blocks to create algorithms.
At the higher level, an object-oriented API simplifies using and combining algorithms.
The next section outlines the rationale behind the structure and design of {\chemtrain} and the API. Then, we provide a more detailed description of the individual modules and implemented functionalities.

\subsection{Software Architecture}
\label{subsec:architecture}

\begin{figure}[htb]
    \centering
    \includegraphics[width=\textwidth]{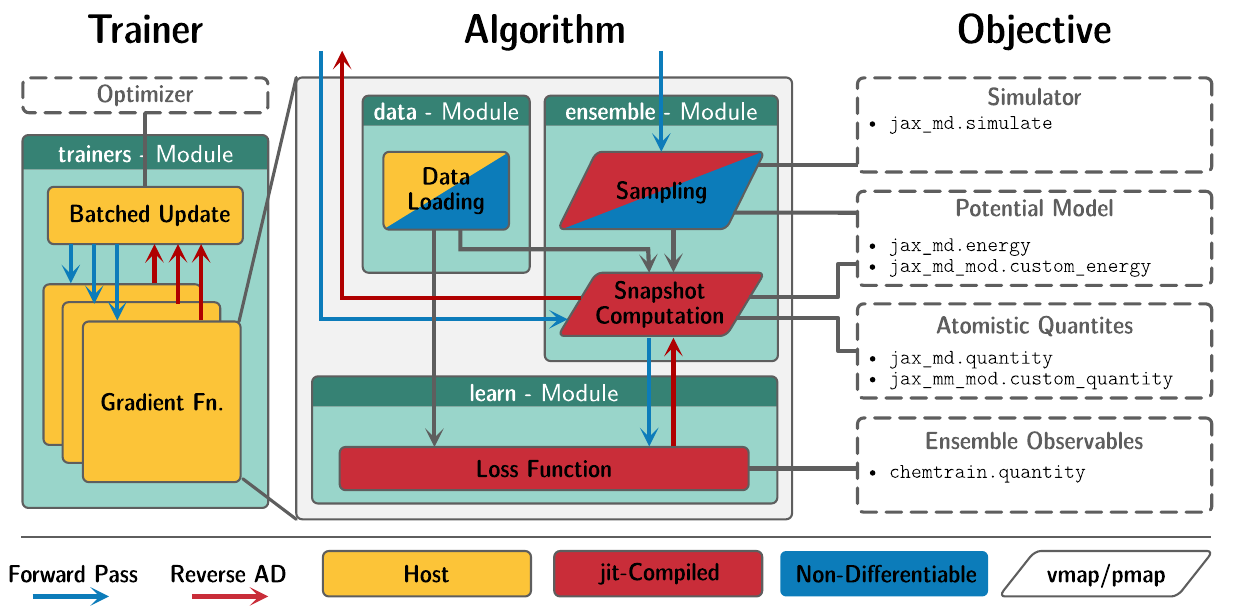}
    \caption{Structure of a Maximum Likelihood Estimate (MLE) learning procedure in {\chemtrain}. The trainer executes algorithms to perform gradient-based optimization of a single or multiple objectives. The algorithms assemble from multi-functional building blocks and rely on a modular definition of the objectives.}
    \label{fig:mle_trainer_template}
\end{figure}

We outline the structure of {\chemtrain} at the example of a Maximum Likelihood Estimation (MLE) algorithm.
Learning a single optimal model maximizing a likelihood or minimizing a loss function, respectively, is the standard scenario for the RM, FM, and DiffTre algorithms (section~\ref{sec:machine_learning}) and easily extensible to an uncertainty-aware perspective, e.g., via the deep-ensemble approach\cite{thalerScalableBayesianUncertainty2023, lakshminarayananSimpleScalablePredictive2017}.
Illustrated in figure~\ref{fig:mle_trainer_template}, learning such a single-point model consists of three components: 
(1) Objectives such as the trainable potential model $U_\theta$, reference systems, thermodynamic state and reference data.
(2) Algorithm to evaluate the loss and loss gradients for the objectives, such as FM for microscopic and DiffTRe for macroscopic reference data.
(3) Trainer, applying a numerical scheme to optimize the model parameters based on the loss.

An important part of {\chemtrain} is efficient and modular implementations of basic training algorithms for the second component.
The FM, RM, and DiffTRe algorithms differ in top-down and bottom-up training perspectives and the predicted quantities.
Nevertheless, the algorithms perform similar routines in the training cycle:
First, the algorithms load reference data (FM/RM) or sample new data from a reference system (RM/DiffTRe).
Based on the sampled or loaded microscopic states, the algorithms compute instantaneous quantities $a(\cdot)$, e.g., forces (FM) and energies (FM/RM/\hspace{0pt}DiffTRe).
Finally, the algorithms compute a loss for the objectives using the reference data and microscopic predictions. 
This loss computation might involve the microscopic predictions directly (FM/RM) or a combination of multiple microscopic predictions, e.g., in the form of weighted ensemble averages (RM/DiffTRe). A computationally efficient implementation of these steps is crucial for the practical performance of the algorithms.
Leveraging the described abstractions, multiple algorithms can benefit from carefully designed implementations.
Therefore, {\chemtrain} follows a modular structure, relying on scalable and flexible building blocks.

The building blocks follow a functional paradigm, relying on a functional implementation of the objectives in the third component.
Therefore, the building blocks can leverage JAX\cite{frostigCompilingMachineLearning} transformations to achieve scalability and flexibility.
A good example of flexibility is the potential model.
The jit-compilation of JAX enables the abstraction of the potential model into an \verb|energy_fn_template|, which creates the model on the fly within the building blocks by applying a parametric architecture to a concrete system.
Listing~\ref{listing:model} provides an example of a $\Delta$-learning potential model using the DimeNet++\cite{gasteigerFastUncertaintyAwareDirectional2022} NN model and a classical force field as prior potential.
The \verb|jit|-compilation enables an efficient evaluation of this on-the-fly-created potential model and further derived quantities, e.g., forces.
Moreover, the vectorization and parallelization capabilities of JAX enable an efficient evaluation of the potential model and other microscopic quantities on multiple parallel samples.
Therefore, {\chemtrain} can scale the computation to use the full power of the devices and distribute workloads among multiple devices.
Finally, the building blocks also benefit from the comprehensive AD functionalities of JAX. 
For example, the FM algorithm can derive a force model directly from the potential model by employing AD.
Moreover, by deriving the parametric force model from the \verb|energy_fn_template|, a second application of AD then yields the gradients of the FM loss function in equation~\eqref{eq:FM_loss}.
Hence, the functional implementation of building blocks enables a simple but scalable definition of objectives.

\begin{lstlisting}[language=Python, caption=Defining a potential model as $\Delta$-learning approach based on a prior potential and an NN potential. {The functional return type enables other building blocks to apply JAX transformations, e.g., computing forces via AD.}, label={listing:model}]
# Setup the graph NN based on the neighbor list
init_fn, gnn_energy_fn = neural_networks.dimenetpp_neighborlist(
    displacement_fn, r_cut, n_species, r_init, nbrs_init,
    embed_size=32, init_kwargs=mlp_init,
)

# Set up a prior potential, e.g., a classical force field
prior_energy_fn = prior_energy(topology, force_field)

# Flexible formulation of trainable potential model
def energy_fn_template(energy_params):

    # Enables obtaining the prior potential from the energy
    # function template
    if energy_params is None:
        return prior_energy_fn
    
    def energy_fn(pos, neighbor, **dynamic):
        # Dynamic quantities can contain the thermostat temperature,
        # e.g., to learn a temperature-dependent CG model
        gnn_energy = gnn_energy_fn(
            energy_params, pos, neighbor, **dynamic
        )

        prior_energy = prior_energy_fn(pos, neighbor=neighbor)
        return prior_energy + gnn_energy
    return energy_fn

\end{lstlisting}

In addition to the modular structure at the lower level, {\chemtrain} provides an object-oriented API to facilitate its application to specialized problems.
This API introduces the concept of trainer classes.
A trainer class wraps single or multiple algorithms, interfacing the problem-specific objectives with the appropriate algorithm and the computation of gradient and loss information with numerical optimization.
The interface between algorithms and objectives simplifies the usage of existing algorithms without exhibiting additional abstraction through modularity.
Therefore, algorithm-specific trainers initialize the building blocks for the algorithms from user-defined objectives.
The interface between numerical optimization and algorithms enables the construction of customized trainers.
For example, alternating the optimization through FM and DiffTRe trainers constitutes an approach to fuse top-down and bottom-up training\cite{rockenAccurateMachineLearning2024}.

\begin{figure}[tbh]
    \centering
    \includegraphics[width=.9\textwidth]{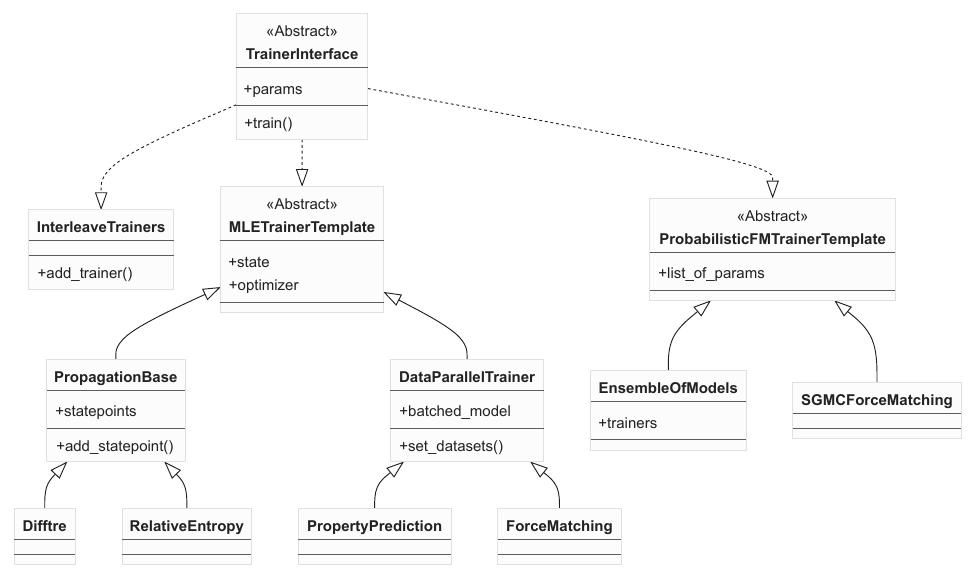}
    \caption{Hierarchy of the trainer classes in {\chemtrain}. The figure displays only a subset of all methods and attributes to illustrate similarities between the trainers.}
    \label{fig:class-hierarchy}
\end{figure}

We implemented the API following an object-oriented paradigm, defining the trainers as classes.
The class-based implementations utilize inheritance to enforce consistency between different trainers and reuse functionalities for similar trainers.
Illustrated in figure~\ref{fig:class-hierarchy}, the class hierarchy in {\chemtrain} reflects the differences between single-point maximum-likelihood estimation and probabilistic learning, e.g., through providing access to a single optimal parameter set (\verb|params|) or multiple samples (\verb|list_of_params|).
Likewise, the hierarchy reflects the respective dependence of DiffTRe and RM on the thermodynamic perturbation formalism, initializing reweighting procedures via the \verb|add_statepoint| method.
Additionally, using stateful trainers provides the advantage that the state of the optimizer, states of the algorithms, and training metrics do not have to be passed explicitly.
In combination with consistency through inheritance, the access to states and metrics through common instance properties simplifies the usage of the algorithms.
Moreover, this combination enables the implementation of more advanced trainers, such as the \verb|InterleaveTrainer| trainer, combining training of FM and DiffTRe by synchronizing the model state (\verb|params|) between successive optimization cycles.

Listing~\ref{listing:difftre_algorithm} illustrates the interplay between the modular structure at the lower level and the object-oriented API in the example of applying the DiffTRe algorithm to two different systems.
The DiffTRe trainer accepts the functional definitions of objectives and optimizer and initializes the states of optimizer and algorithms.
Due to the clear separation between data and computation, individual definitions, e.g., the \verb|energy_fn_template|, are reusable for different standpoints or problem definitions.
Since DiffTRe inherits from the \verb|MLEtrainer| abstract base class like other MLE algorithms (see figure~\ref{fig:class-hierarchy}), training is possible through common methods such as \verb|trainer.train(...)| and \verb|trainer.save_energy_params(...)|.

\begin{lstlisting}[language=Python, caption=Initializing the DiffTRe algorithm., label={listing:difftre_algorithm}]
# All statepoints share the same parameters
trainer = trainers.Difftre(init_params, optimizer)

# Multiple statepoints can be added, representing a specific
# system at a fixed thermodynamic state-point
trainer.add_statepoint(
    energy_fn_template, simulator_template, neighbor_fn, timings,
    state_kwargs, snapshot_fns, reference_state, targets, observable_fns)
trainer.add_statepoint(
    energy_fn_template, ...
)

# Perform gradient-based training
trainer.train(num_updates)

# Save the trained energy parameters via pickle
trainer.save_energy_params('trainer.pkl', save_format='.pkl')

# Access training metrics, such as predictions, loss, ...
... = trainer.predictions
... = trainer.batch_losses
\end{lstlisting}

\subsection{Core Modules}

In the following section, we provide a closer description of the modules of {\chemtrain}.
Table~\ref{tab:building_blocks} contains an overview of the most important building blocks arranged into the core modules of {\chemtrain}.

\begin{table}[ht]
    \centering
    \begin{tabular}{l| p{.8\textwidth}}
         \textbf{Module} & \textbf{Content} \\\hline\hline
          \texttt{data} & \textbf{Data Loading:} Loading from \texttt{numpy} with shuffling/splitting/subsampling, initializing of JaxSGMC \texttt{DataLoaders} \cite{thalerJaxSGMCModularStochastic2024}\newline
          \textbf{Data Processing:} Scaling to fractional coordinates, linear CG mapping from Multiscale CG method\cite{noidMultiscaleCoarsegrainingMethod2008}\\\hline
          
          \texttt{ensemble} & 
          \textbf{Reweighting:} Reference/Adaptive reweighting (NVT), BAR method\cite{bennett1976, wyczalkowski2010} (NVT)
          \newline
          \textbf{Simulation:} (Batched) execution of JAX, M.D.\cite{schoenholzJAXFrameworkDifferentiable2021} simulations with equilibration/subsampling\newline
          \textbf{Evaluation:} batched snapshot computation 
          \newline \textbf{Utilities:} Initialization of simulator template
          \\\hline

          \texttt{quantity} & \textbf{Learnable Quantities:} dipole moment, partial charges\cite{thalerActiveLearningGraph2024}
          \newline
          \textbf{Macrosopic Observables:} (linearized\cite{imbalzano2020}) ensemble average, heat capacities\cite{sadusMolecularSimulationOrthobaric2019, stroker2021}, relative entropy
          \newline
          \textbf{Utilities:} Helper functions to initialize DiffTRe targets\\\hline
          
          \texttt{learn} & \textbf{Algorithms:} DiffTRe\cite{thalerLearningNeuralNetwork2021}, (probabilistic) Force Matching\cite{thalerDeepCoarsegrainedPotentials2022, thalerScalableBayesianUncertainty2023}, Relative Entropy Minimization\cite{thalerDeepCoarsegrainedPotentials2022}
          \newline
          \textbf{Utilities:} \texttt{pmap}/\texttt{shmap}-parallelization of loss/validation function, masked MSE/MAE functions\\\hline 
          
          \texttt{trainers} & \textbf{Core Trainers:} DiffTRe\cite{thalerLearningNeuralNetwork2021}, PropertyPrediction\cite{thalerActiveLearningGraph2024}, ForceMatching\cite{thalerDeepCoarsegrainedPotentials2022}, RelativeEntropy\cite{thalerDeepCoarsegrainedPotentials2022}, SGMCForceMatching\cite{thalerScalableBayesianUncertainty2023}
          \newline
          \textbf{Advanced Trainers:} EnsembleOfModels, InterleaveTrainers\\\hline
          
    \end{tabular}
    \caption{Overview of the most important building blocks of {\chemtrain}.}
    \label{tab:building_blocks}
\end{table}

\subsubsection{{\tt data} -- Loading and Processing of Reference Data}

{\chemtrain} learns potential models via data-driven approaches from experimental or simulated reference data.
In the bottom-up approach, the reference data consists of microscopic observations with typically higher spatial and temporal resolution than the macroscopic reference data used in top-down learning.
Hence, efficient data management and stochastic optimization based on subsamples of the data are crucial to ensure the scalability of the bottom-up approaches.

In the \verb|data.data_loaders| module, {\chemtrain} provides an interface to JaxSGMC\cite{thalerJaxSGMCModularStochastic2024}.
At its current state, JaxSGMC simplifies the usage of reference data in the \verb|numpy| format, supporting multiple batching strategies and loading of data into \verb|jit|-compiled functions.
Moreover, following a modular concept like {\chemtrain}, JaxSGMC enables setting up custom data-loading pipelines, e.g., to load reference data from other formats such as HDF5\cite{hdf5}.
Therefore, the \verb|data.data_loaders| module ensures the extensibility of {\chemtrain} to other data sources and data management strategies.

Additionally, the \verb|data.preprocessing| module provides routines to load and pre-process reference data in the \verb|numpy| format.
These routines include typical tasks for data-driven learning, such as shuffling, subsampling, and data splitting.
Moreover, the module provides preprocessing functionalities more specific to learning potential models, such as a CG mapping routine complying with the Multiscale CG method\cite{noidMultiscaleCoarsegrainingMethod2008} (FM, section~\ref{subsubsec:force_matching}).

\subsubsection{{\tt ensemble} -- Simulating and Importance Sampling Ensembles}

The \verb|ensemble| module complements the \verb|data| module by providing utilities to generate data.
In detail, the \verb|ensemble| module divides into three submodules.
First, the submodule \verb|ensemble.sampling| provides routines to sample new states $\bm r$ from an ensemble $p_U(\bm r)$, e.g., for DiffTRe and RM.
Second, the submodule \verb|ensemble.evaluation| contains functions to evaluate microscopic quantities $a(\bm r)$ for large numbers of samples $\bm r$.
Finally, the \verb|ensemble.reweighting| module combines both functionalities to provide variants of thermodynamic perturbation theory.

In many cases, MD is a practical method to draw samples from an ensemble of molecular systems (section~\ref{section:theory}).
Therefore, the \verb|ensemble.sampling| module interfaces {\chemtrain} to the MD integrators of JAX, M.D.\cite{schoenholzJAXFrameworkDifferentiable2021} providing several classical integrators to sample from various ensembles, e.g., NVE, NVT, and NPT.
The \verb|ensemble.sampling| module combines these integrators with the flexible \verb|energy_fn_template| (listing~\ref{listing:model}) to obtain a parametric MD simulation routine.
This routine follows a typical simulation procedure, discarding states during an equilibration cycle and keeping decorrelated states during the production cycle via subsampling.
Moreover, the routine enables simulations at constant or time-dependent thermodynamic states, passing the properties of the current state to the energy function via keyword arguments (\verb|**dynamic|, listing~\ref{listing:model}).
Additionally, through a fully functional implementation, the simulation routine can leverage {JAX} \verb|vmap| to accelerate phase-space exploration by simulating multiple systems in parallel.
In principle, the interface also supports advanced simulation methods such as tempered/meta-dynamics replica exchange\cite{sugitaReplicaexchangeMolecularDynamics1999, gil-leyEnhancedConformationalSampling2015}.

A crucial part of top-down and bottom-up training algorithms is the computation of microscopic quantities $a(\bm r)$ for many sampled conformations $\bm r$.
The submodule \verb|ensemble.evaluation| provides functions to perform this computation efficiently by performing automatically batched computations via \verb|vmap|.
In line with the modular concept of {\chemtrain}, the implementation of the microscopic quantities follows a simple but flexible interface, displayed in listing~\ref{listing:snapshot}.
Derived from the states of the JAX, M.D. simulators, the \verb|state| argument bundles the internal state of a system.
Essentially, this state contains the particle positions $\bm r$ but can also include additional properties, such as the box size for NPT ensembles.
The keyword arguments provide external parameters, e.g., the thermostat pressure, temperature, and learnable parameters $\theta$.
Moreover, the keyword arguments can provide pre-extracted features, such as the neighbor list, shared among the microscopic quantities for computational efficiency.
Additional pre-extracted features are simple to specify, following the same protocol as the microscopic quantities.

\begin{lstlisting}[language=Python, caption=Atomistic (snapshot) quantity, label={listing:snapshot}]
def compute_fn(state, neighbor=None, **kwargs) -> PyTree:
    # Computation, e.g., RDF or pressure
    ...
    return quantity_snapshot
\end{lstlisting}

The submodule \verb|ensemble.reweighting| implements different variants of thermodynamic perturbation theory\cite{zwanzig2004}.
Combined with the \verb|sampling| and \verb|evaluation| submodules, one variant is a streamlined version of umbrella sampling\cite{torrie1977} for NVT and NPT ensembles with automatic re-sampling of reference states $\bm r$.
Crucial for the DiffTRe and RM algorithms, this umbrella sampling variant provides an efficient end-to-end differentiable routine to compute macroscopic equilibrium properties (section~\ref{subsec:top_down}).
Moreover, by employing the BAR algorithm\cite{bennett1976, wyczalkowski2010} and Zwanzig's relation~\cite{zwanzig2004} between successive resampling steps, this umbrella sampling variant estimates the change of free energy $\Delta Q$ and entropy $\Delta S$ between the initial and current potential $U_\theta$ on-the-fly\cite{rockenPredictingSolvationFree2024}. 
Additional routines enable reweighting based on pre-computed data and between different thermodynamic state points.

\subsubsection{{\tt quantity} -- Microscopic and Macroscopic Observables}

The DiffTRe algorithm provides a method to learn macroscopic equilibrium properties $\mathcal A$ (section~\ref{subsec:top_down}).
Therefore, the \verb|quantity.observables| module contains implementations of $\mathcal A$ from microscopic properties $a(\bm r)$ compatible with the underlying thermodynamic perturbation theory.
In the simplest scenario, the macroscopic properties are weighted ensemble averages of a single type of microscopic properties, e.g., the instantaneous radial distribution function or microscopic pressure.
However, some thermodynamic quantities, such as the isobaric heat capacity\cite{stroker2021}, depend on multiple ensemble averages formed by multiple microscopic properties.
To flexibly support such more complicated observables, {\chemtrain} passes all microscopic quantities and the current weights to the compute functions of the macroscopic observables.
Listing~\ref{listing:observable} provides an example implementation of an observable, computing the variance $\operatorname{Var}_{U}[a] = {\langle a^2 \rangle_U - \langle a \rangle_U^2}$ of an instantaneous observable $a(\bm r)$.
Due to passing all computed microscopic predictions, the logic of selecting the required quantities remains within the compute function for the macroscopic observable, and microscopic predictions are only computed once.
Moreover, the computation of more involved observables, e.g., containing the variance of a microscopic property, is possible.
Nevertheless, the separation between microscopic and macroscopic observables can complicate the usage.
Therefore, the additional \verb|quantity.targets| module aims to mitigate this complication by providing routines to set up macroscopic observables alongside required microscopic observables for common quantities, such as the radial and angular distribution functions or the pressure.

\begin{lstlisting}[language=Python, caption=Example implementation of the weighted ensemble average., label={listing:observable}]
def obs_variance(snapshots, weights=None, **statepoint_kwargs):
    # A dynamic definition of the statpoint is possible
    del statepoint_kwargs # Unused by computation

    # In general, the compute function has access to all snapshots
    snapshot = snapshots['quantity_key']
    snapshot_sq = snapshot ** 2

    # Weights are only necessary for the DiffTRe algorithm
    if weights is not None:
        weights *= weights.size
        snapshot = (snapshot.T * weights).T
        snapshot_sq = (snapshot_sq.T * weights).T

    # Variance is Var[x] = E[xx] - E[x]E[x]
    var = jnp.mean(snapshot_sq, axis=0)
    var -= jnp.mean(snapshot, axis=0) ** 2
        
return var
\end{lstlisting}

In addition to modeling complicated potential energy surfaces, NNs can also accurately model other molecular or atomistic properties, such as partial charges.
Therefore, the \verb|quantity.property_prediction|  module extends the classical microscopic quantities, e.g., correlation functions or pressures, with learnable quantities such as partial charges\cite{thalerActiveLearningGraph2024}.
In detail, the submodule provides tools to derive property predictions from NNs, interfacing these predictors with the \verb|compute_fn| protocol (listing~\ref{listing:snapshot}) expected by the \verb|ensemble.evaluation| module.

\subsubsection{{\tt learn} --- (Deep) Learning Algorithms}

The \verb|learn| module contributes building blocks for standard training algorithms.
These building blocks comprise general utilities and concrete routines for the core algorithms of {\chemtrain}.
For example, {the module provides general utilities that evaluate the model, compute the loss,} and perform the gradient update on multiple devices using JAX \verb|pmap| and \verb|shmap| transformations.
{Trainers can apply such general utilities or other JAX transformations to functional implementations of larger algorithm parts, i.e., concrete routines, to efficiently employ algorithms.}
Therefore, {the} \verb|data| module contains the highest functional layer of the FM, RM, and DiffTRe algorithms.

\subsection{{\tt trainers} --- High-Level API to Algorithms}

The \verb|trainers| module contains the high level API of {\chemtrain}.
Besides the wrappers around the core algorithms described in section~\ref{subsec:architecture}, this API enables the creation of more advanced training routines.
For example, listing~\ref{listing:interleaved} demonstrates the fusion of the FM and DiffTRe algorithm through the higher-level \verb|InterleaveTrainers| trainer.
Additionally, the \verb|trainers.extensions| submodule provides plugin-functionalities to extend routines of the trainers, e.g., to enhance their usability.
For example, the submodule provides a decorator to the DiffTRe and FM trainers to log the training statistics to the ML framework Weights~\&~Biases\cite{wandb} after each epoch.
Therefore, the trainers API provides much flexibility to extend the functionality and usability of the core training algorithms without the need to make difficult changes to the core of the {\chemtrain} library.

\begin{lstlisting}[language=Python, caption=Combining multiple trainers., label={listing:interleaved}]
# Allows combining the trainers, executing them in the order in which
# they were added
trainer = trainers.InterleaveTrainers(full_checkpoint=False)

# Run force matching 10 epochs before running difftre for 2 epochs
trainer.add_trainer(fm_trainer, num_updates=10, name='Force Matching')
trainer.add_trainer(difftre_trainer, num_updates=2, name='DiffTRe')

# Training as usual for MLE trainers
trainer.train(100, checkpoint_frequency=10)
\end{lstlisting}

\section{Examples}
\label{section:examples}

In the following section, we illustrate the usage of {\chemtrain} in two examples.
In the first example, we combine the bottom-up approaches FM and RM to efficiently learn an accurate structural representation of a heavy-atom alanine dipeptide model in implicit water\cite{thalerDeepCoarsegrainedPotentials2022}.
In the second example, we combine bottom-up and top-down learning via the FM and DiffTRe algorithms to train an atomistic titanium model consistent with experimental measurements of thermodynamic properties\cite{rockenAccurateMachineLearning2024}.
The examples briefly demonstrate the importance of combining multiple learning strategies to efficiently develop accurate potential models.
We refer to the original papers and the reference documentation for detailed descriptions {and results}.

\subsection{Bottom-Up CG Model of Alanine Dipeptide in Implicit Water}

Alanine Dipeptide is a classical benchmark system for CG training methods to test their ability to model free energy surfaces with multiple meta-stable states. The reference data and training setup follow the original paper\cite{thalerDeepCoarsegrainedPotentials2022}.
In short, the implicit solvent CG model consists of the 10 heavy atoms of Alanine Dipeptide, without hydrogen atoms or water molecules. The reference data was derived from a $100$~ns atomistic reference trajectory at $T_\mathrm{ref} = 300$~K.
Building on the $\Delta$-learning ansatz \cite{Ramakrishnan2015}, the potential model consists of a learnable correction $\Delta U(\theta, \mathbf{R})$ via the DimeNet++\cite{gasteigerFastUncertaintyAwareDirectional2022, thalerLearningNeuralNetwork2021} message-passing neural network and a prior potential
\begin{linenomath}\begin{equation}
    \label{eq:ala_prior}
    U^\mathrm{Prior}(\mathbf{R}) = \sum_{i} U^H_i(b_i) + \sum_{j} U^H_j(\alpha_j) + \sum_{k} U^D_k(\omega_k)
\end{equation}\end{linenomath}
based on on harmonic terms $U^H$ for all bonds $b_i$ and angles $\alpha_j$ and cosine series potentials $U^D_i$ for the proper dihedral angles of the backbone $\omega_i$.

Listing~\ref{listing:alanine} illustrates the training setup in {\chemtrain}.
First, we set up the prior potential by defining the parameters in a \texttt{toml} file and generating the topology from the initial structure.
After defining the complete potential following listing~\ref{listing:model}, we train the model via FM for $100$ epochs.
Finally, we set up a Langevin thermostat to sample from the learned ensemble and fine-tune the model by RM for $50$ updates.

\begin{lstlisting}[language=Python, caption=Training a $\Delta$-learning model following listing~\ref{listing:model} for Alanine Dipeptide with force matching and relative entropy minimization., label={listing:alanine}]
# Define prior potential
force_field = prior.ForceField.load_ff("../_data/alanine_heavy.toml")
top = mdtraj.load_topology("../_data/alanine_heavy_2_7nm.gro")
topology = prior.Topology.from_mdtraj(top, mapping)

# setup total model
def energy_fn_template(energy_params): ...

# Set up and train with force matching
force_matching = trainers.ForceMatching(
    init_params, fm_optimizer, energy_fn_template, nbrs_init,
    batch_per_device=500 // len(jax.devices())
)
force_matching.set_datasets({
    'R': position_dataset[:5e5, ...], 
    'F': force_dataset[:5e5, ...]
    }, train_ratio=0.7 
)
force_matching.train(100)
force_matching.save_energy_params("alanine_dipeptide_fm_params.pkl", '.pkl')
force_matching.save_trainer("alanine_dipeptide_fm_trainer.pkl", '.pkl')

# Define simulator, RM trainer, add simulation and fine-tune model
init_ref_state, sim_template = sampling.initialize_simulator_template(
    simulate.nvt_langevin, shift_fn=shift_fn, nbrs=nbrs_init,
    extra_simulator_kwargs={"kT": 300. * quantity.kb, "gamma": 100, "dt": 0.002}
)
rm_post_fm = trainers.RelativeEntropy(
    force_matching_params, optimizer, reweight_ratio=1.1,
    energy_fn_template=energy_fn_template)
rm_post_fm.add_statepoint(
    position_dataset[:4e5, ...],
    energy_fn_template, sim_template, neighbor_fn,
    re_timings, state_kwargs, reference_state,
    reference_batch_size=4e5,
    vmap_batch=100, resample_simstates=True)
relative_entropy.train(50)
\end{lstlisting}

\begin{figure}
    \centering
    \includegraphics[width=\textwidth]{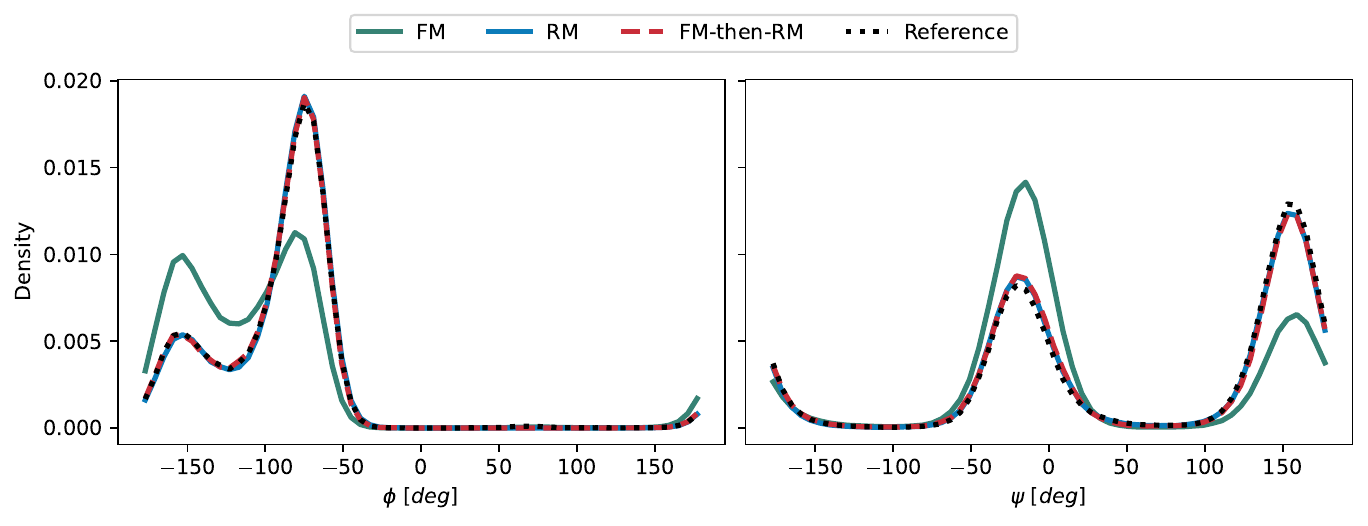}
    \caption{Density histograms of the backbone dihedral angels of alanine dipeptide from simulations with Force Matching (FM), Relative Entropy Minimization (RM), and FM-then-RM trained models in comparison to an atomistic reference simulation.}
    \label{fig:alanine_FES}
\end{figure}

After training, we compare the FM model to an RM model trained for 300 updates and the FM model fine-tuned via 50 RM updates.
Therefore, we sample 100 ns worth of MD simulations for each model and compare the obtained dihedral angle distribution to the atomistic reference (figure \ref{fig:alanine_FES}). 
The FM model deviates {more} strongly from the reference density {than the RM model}. {This performance difference to the RM model can arise from the finite model capacity that affects the optimal FM solution\cite{chaimovichCoarsegrainingErrorsNumerical2011} and errors from the MD simulation\cite{thalerDeepCoarsegrainedPotentials2022}. Moreover, RM converges evidently to its optimal potential with substantially fewer data\cite{thalerDeepCoarsegrainedPotentials2022, kohlerFlowMatchingEfficientCoarseGraining2023}.}
However, when {retraining} the FM model via 50 RM updates, the model matches the atomistic reference as accurately as trained solely via RM for 300 updates.
We attribute the improved performance of the hybrid training largely to the improved initial model state provided by FM.
{As the targets of FM and RM partially align\cite{thalerDeepCoarsegrainedPotentials2022}, performing RM from the FM model rather than a random initialization circumvents traversing a larger distance in parameter space. Therefore, RM can directly fine-tune a more suitable initial model, reducing computational demands.}

\subsection{Atomistic Model of Titanium via Experimental and Simulation Data Fusion}

This example illustrates the importance of fusing experimental and simulation reference data to obtain an atomistic model of titanium consistent with ab initio computations and experimental measurements.
The training procedure and reference data follow the paper by Röcken and Zavadlav~\cite{rockenAccurateMachineLearning2024}.
From the original paper, we take the DimeNet++ model pre-trained on a curated dataset of energies, forces, and virials computed via density functional theory for various conformations of bulk titanium~\cite{wen2021specialising}. Additionally, we take the experimental measurements of bulk elastic constants at multiple temperatures~\cite{simmons1971single}.

Illustrated in listing~\ref{listing:AT_titanium}, we employ FM on the ab initio data and DiffTRe on the experimental data at $323$ and $923$~K to fuse microscopic and macroscopic reference data in {\chemtrain}.
We first set up both trainers separately, defining the computations of microscopic quantities for FM and DiffTRe and the computations of macroscopic observables only for DiffTRe. Then, we leverage the \verb|InterleaveTrainers| trainer for 100 repetitions, each sequentially training the model for one epoch of FM and one update for all statepoints of DiffTRe.
\begin{lstlisting}[language=Python, caption=Fused bottom-up and top-down learning of an atomistic NN model of titanium., label={listing:AT_titanium}]
# Initialize an FM trainer with pre-trained params
trainer_fm = trainers.ForceMatching(
    pretrained_params, optimizer_fm, energy_fn_template, nbrs_init,
    batch_per_device=30, batch_cache=8,
    gammas={'W': 4e-6,'U': 1e-6, 'F': 1e-2},
    additional_targets={'W': virial_snapshot_fn},
    weights_keys={'W': 'W_weights'}
)
# Load training data analogous to the previous example
trainer_fm.set_datasets(...)

# Define the observables for top-down training, the required snapshots,
# and the simulator to sample states
observables = {
    'C_const': observables.init_born_stiffness_tensor(...),
    'pressure': observables.init_traj_mean_fn('pressure')
}
compute_fns = {
    'born_stiffness': born_stiffness_snapshot_fn,
    'born_stress': born_stress_snapshot_fn,
    'pressure': pressure_snapshot_fn
}
# Analogous to previous example
init_ref_state, sim_template = sampling.initialize_simulator_template(...)

# Initialize DiffTRe to train elastic constants and pressure
trainer_difftre = trainers.Difftre(
    init_params, optimizer_difftre, sim_batch_size=-1, reweight_ratio=1.0
)

# Add a separate statepoint for each temperature. We can re-use the
# snapshot and observable functions for all statepoints
for temp in [323, 923]:
    targets = {'C_const': {'target': exp_data[temp], 'gamma': 1e-10},
               'pressure': {'target': 1. / 16.6054, 'gamma': 1e-9}}}
    reference_state = init_ref_state(...) # Analogous to previous example
    trainer_difftre.add_statepoint(
        energy_fn_template, sim_template, neighbor_fn_difftre, timings,
        state_kwargs=state_kwargs[temp], reference_state=reference_state,
        targets=targets, observables=observables, quantities=compute_fns)

# Optimize sequentially on both objectives
trainer_fused = trainers.InterleaveTrainers(sequential=True)
trainer_fused.add_trainer(trainer_fm, num_updates=1, name='FM')
trainer_fused.add_trainer(trainer_difftre, num_updates=1, name='DiffTRe')
trainer_fused.train(100)
\end{lstlisting}

\begin{figure}[!htb]
    \centering
    \includegraphics[width=\textwidth]{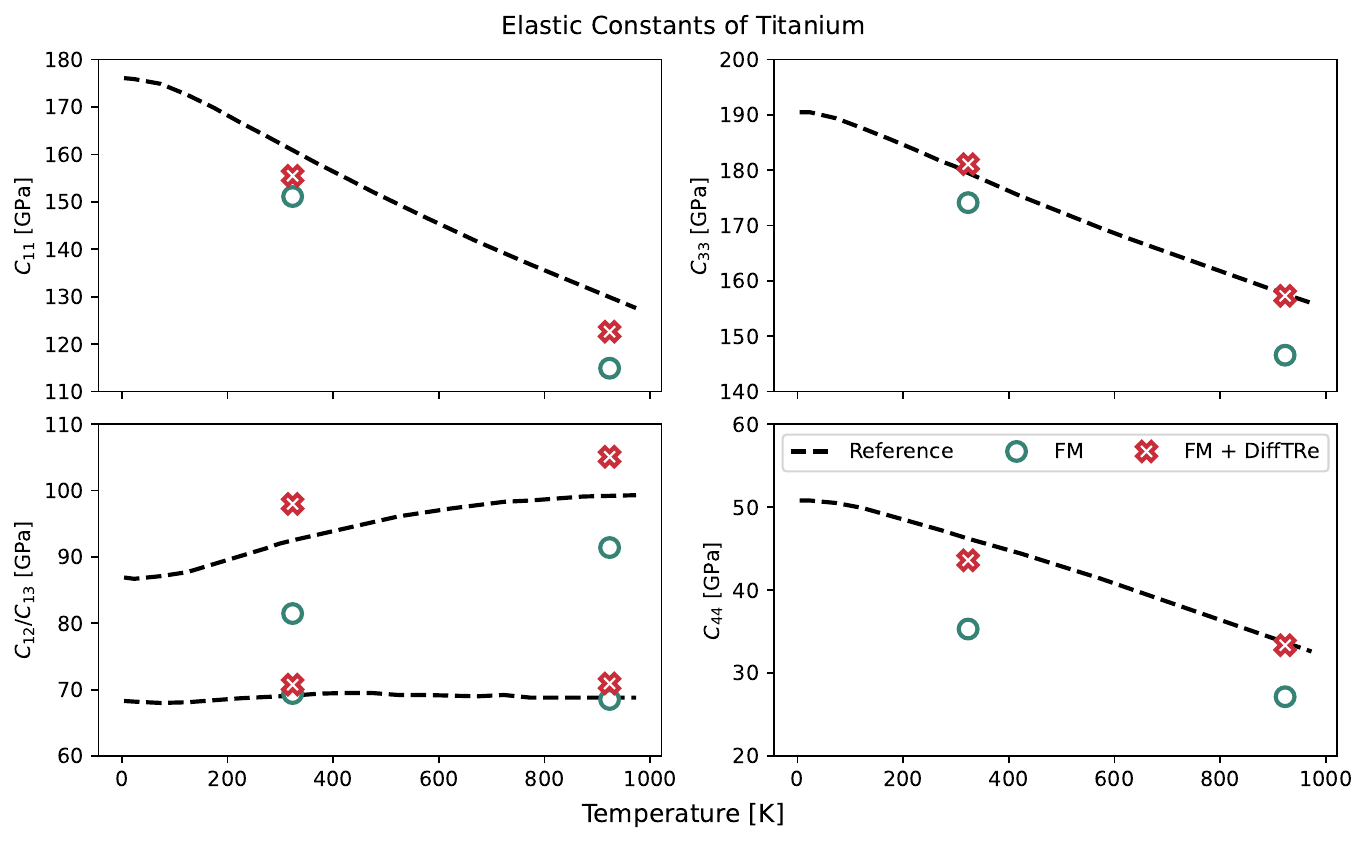}
    \caption{Predicted elastic constants. The black dashed line denotes the experimental reference, the green circles denote the prediction of the bottom-up trained model via Force Matching (FM), and the red stars denote the predictions after sequentially training bottom-up with FM and top-down with Differential Trajectory Reweighting (DiffTRe).}
    \label{fig:Ti_EC}
\end{figure}

To compare both models, we compute the elastic constants in a $70\ \mathrm{ps}$ long reference simulation withholding $10\ \mathrm{ps}$ of equilibration.
Figure~\ref{fig:Ti_EC} shows that the fusion of top-down and bottom-up learning can improve the agreement of the model and experiments.
We expect further improvement by training for more epochs and using a longer sampling interval than $20\ \mathrm{ps}$ for training with DiffTRe.

\section{Conclusion}
\label{sec:conclusion}

In this work, we presented {\chemtrain}, a framework that enables the training of sophisticated NN potentials through customizable training routines. Similarities between commonly applied top-down and bottom-up training procedures allow the decomposition of the algorithms into standard building blocks. Therefore, we implemented {\chemtrain} in a modular structure, following a functional paradigm at the lower level to leverage JAX transformations to obtain scalable and reusable building blocks. An object-oriented API built on the modular structure enables a simple combination or extension of algorithms.
 
Our experiments demonstrate that combining algorithms as performed for classical force fields is also beneficial for training NN potential models. 
On the one hand, combining methods enables a pre-training of the potential with less expensive methods to reduce the cost of more computationally demanding but more accurate or data-efficient methods.
On the other hand, supplementing bottom-up methods with top-down approaches expands the amount of usable data and enforces predictions consistent with experiments.
Therefore, we expect that {\chemtrain} can contribute to expediting the success of NN potential models for molecular simulations.

For the future, we plan to extend {\chemtrain} by providing further advanced training routines. For example, active learning can improve data efficiency \cite{smithApproachingCoupledCluster2019} and is, in principle, straightforward to add due to the flexible API of {\chemtrain}. Additionally, we plan to extend the supported macroscopic observables, such as dynamical observables and better free energy perturbation estimates, by generalizing the DiffTRe algorithm. Moreover, we aim to accelerate the DiffTRe algorithm by integrating learnable advanced sampling schemes with {\chemtrain}.
As it is unlikely that JAX, M.D. will catch up with established MD packages regarding robustness, flexibility, and computational efficiency in the foreseeable future, we also intend to tighten the integration of {\chemtrain} with classical MD software\cite{thompsonLAMMPSFlexibleSimulation2022, abrahamGROMACSHighPerformance2015, andersonGeneralPurposeMolecular2008,andersonHOOMDbluePythonPackage2020} to simplify the deployment of the trained NN potentials to large scale simulations.

\section*{Author Contributions}
\textbf{Paul Fuchs}: Methodology, Software, Validation, Formal analysis, Investigation, Writing - Original Draft, Visualization
\textbf{Stephan Thaler}: Conceptualization, Methodology, Software, Writing - Original Draft, Writing - Review $\&$ Editing
\textbf{Sebastien Röcken}: Methodology, Software, Validation, Formal analysis, Investigation, Writing - Original Draft, Visualization
\textbf{Julija Zavadlav}: Conceptualization, Resources, Writing - Review $\&$ Editing, Supervision, Project administration, Funding acquisition.

\section*{Acknowledgments}
This work was funded by the Deutsche Forschungsgemeinschaft (DFG, German Research Foundation) - 534045056.

Funded by the European Union. Views and opinions expressed are however those of the author(s) only and do not necessarily reflect those of the European Union or the
European Research Council Executive Agency. Neither the European Union nor the granting
authority can be held responsible for them. Funded by the European Research Council (ERC) StG under Grant No. 101077842—SupraModel.


\bibliographystyle{elsarticle-num}
\bibliography{references}

\end{document}